\documentclass[aps,prl,twocolumn,superscriptaddress,nofootinbib,floatfix]{revtex4-2}

\usepackage{graphicx}

\usepackage{xcolor} % To be removed

\begin{document}

\title{First Evidence of Axial Shape Asymmetry and Configuration Coexistence in $^{74}$Zn: \\ Suggestion for a Northern Extension of the $N=40$ Island of Inversion}

\author{M.~Rocchini}
\email{mrocchin@uoguelph.ca}
\affiliation{Department of Physics, University of Guelph, N1G 2W1 Guelph, Canada}

\author{P.E.~Garrett}
\affiliation{Department of Physics, University of Guelph, N1G 2W1 Guelph, Canada}

\author{M.~Zieli\'nska}
\affiliation{IRFU, CEA, Universit\'e Paris-Saclay, F-91191 Gif-sur-Yvette, France}

\author{S.M.~Lenzi}
\affiliation{Dipartimento di Fisica, Universit\`a di Padova, I-35122 Padova, Italy}
\affiliation{INFN Sezione di Padova, I-35131 Padova, Italy}

\author{D.D.~Dao}
\affiliation{Universit\'e de Strasbourg, CNRS, IPHC UMR 7178, F-67000 Strasbourg, France}

\author{F.~Nowacki}
\affiliation{Universit\'e de Strasbourg, CNRS, IPHC UMR 7178, F-67000 Strasbourg, France}

\author{V.~Bildstein}
\affiliation{Department of Physics, University of Guelph, N1G 2W1 Guelph, Canada}

\author{A.D.~MacLean}
\affiliation{Department of Physics, University of Guelph, N1G 2W1 Guelph, Canada}

\author{B.~Olaizola}
\altaffiliation[Present address: ]{CERN, CH-1211 Geneva, Switzerland.}
\affiliation{TRIUMF, V6T 2A3 Vancouver, Canada}

\author{Z.T.~Ahmed}
\affiliation{Department of Physics, University of Guelph, N1G 2W1 Guelph, Canada}

\author{C.~Andreoiu}
\affiliation{Department of Chemistry, Simon Fraser University, V5A 1S6 Burnaby, Canada}

\author{A.~Babu}
\affiliation{TRIUMF, V6T 2A3 Vancouver, Canada}

\author{G.C.~Ball}
\affiliation{TRIUMF, V6T 2A3 Vancouver, Canada}

\author{S.S.~Bhattacharjee}
\altaffiliation[Present address: ]{Institute of Experimental and Applied Physics, Czech Technical University in Prague, 110 00 Prague, Czech Republic.}
\affiliation{TRIUMF, V6T 2A3 Vancouver, Canada}

\author{H.~Bidaman}
\affiliation{Department of Physics, University of Guelph, N1G 2W1 Guelph, Canada}

\author{C.~Cheng}
\affiliation{TRIUMF, V6T 2A3 Vancouver, Canada}

\author{R.~Coleman}
\affiliation{Department of Physics, University of Guelph, N1G 2W1 Guelph, Canada}

\author{I.~Dillmann}
\affiliation{TRIUMF, V6T 2A3 Vancouver, Canada}
\affiliation{Department of Physics and Astronomy, University of Victoria, V8P 5C2 Victoria, Canada}

\author{A.B.~Garnsworthy}
\affiliation{TRIUMF, V6T 2A3 Vancouver, Canada}

\author{S.~Gillespie}
\affiliation{TRIUMF, V6T 2A3 Vancouver, Canada}

\author{C.J.~Griffin}
\affiliation{TRIUMF, V6T 2A3 Vancouver, Canada}

\author{G.F.~Grinyer}
\affiliation{Department of Physics, University of Regina, S4S 0A2 Regina, Canada}

\author{G.~Hackman}
\affiliation{TRIUMF, V6T 2A3 Vancouver, Canada}

\author{M.~Hanley}
\affiliation{Department of Physics, Colorado School of Mines, CO 80401 Golden, USA}

\author{A.~Illana}
\affiliation{Accelerator Laboratory, Department of Physics, University of Jyv\"askyl\"a, FI-40014 Jyv\"askyl\"a, Finland}

\author{S.~Jones}
\affiliation{Department of Physics and Astronomy, University of Tennessee, TN 37996 Knoxville, USA}

\author{A.T.~Laffoley}
\affiliation{Department of Physics, University of Guelph, N1G 2W1 Guelph, Canada}

\author{K.G.~Leach}
\affiliation{Department of Physics, Colorado School of Mines, CO 80401 Golden, USA}

\author{R.S.~Lubna}
\altaffiliation[Present address: ]{Facility for Rare Isotope Beams, Michigan State University, MI 48824 East Lansing, USA.}
\affiliation{TRIUMF, V6T 2A3 Vancouver, Canada}

\author{J.~McAfee}
\affiliation{TRIUMF, V6T 2A3 Vancouver, Canada}
\affiliation{Department of Physics, University of Surrey, GU2 7XH Guildford, UK}

\author{C.~Natzke}
\affiliation{TRIUMF, V6T 2A3 Vancouver, Canada}
\affiliation{Department of Physics, Colorado School of Mines, CO 80401 Golden, USA}

\author{S.~Pannu}
\affiliation{Department of Physics, University of Guelph, N1G 2W1 Guelph, Canada}

\author{C.~Paxman}
\affiliation{TRIUMF, V6T 2A3 Vancouver, Canada}
\affiliation{Department of Physics, University of Surrey, GU2 7XH Guildford, UK}

\author{C.~Porzio}
\altaffiliation[Present address: ]{Lawrence Berkeley National Laboratory, CA 94720 Berkeley, USA.}
\affiliation{TRIUMF, V6T 2A3 Vancouver, Canada}
\affiliation{INFN Sezione di Milano, I-20133 Milano, Italy}
\affiliation{Dipartimento di Fisica, Universit\`a di Milano, I-20133 Milano, Italy}

\author{A.J.~Radich}
\affiliation{Department of Physics, University of Guelph, N1G 2W1 Guelph, Canada}

\author{M.M.~Rajabali}
\affiliation{Physics Department, Tennessee Technological University, Cookeville, Tennessee 38505, USA}

\author{F.~Sarazin}
\affiliation{Department of Physics, Colorado School of Mines, CO 80401 Golden, USA}

\author{K.~Schwarz}
\affiliation{TRIUMF, V6T 2A3 Vancouver, Canada}

\author{S.~Shadrick}
\affiliation{Department of Physics, Colorado School of Mines, CO 80401 Golden, USA}

\author{S.~Sharma}
\affiliation{Department of Physics, University of Regina, S4S 0A2 Regina, Canada}

\author{J.~Suh}
\affiliation{Department of Physics, University of Regina, S4S 0A2 Regina, Canada}

\author{C.E.~Svensson}
\affiliation{Department of Physics, University of Guelph, N1G 2W1 Guelph, Canada}

\author{D.~Yates}
\affiliation{TRIUMF, V6T 2A3 Vancouver, Canada}
\affiliation{Department of Physics and Astronomy, University of British Columbia, V6T 1Z4 Vancouver, Canada}

\author{T.~Zidar}
\affiliation{Department of Physics, University of Guelph, N1G 2W1 Guelph, Canada}

\date{\today}

\begin{abstract}

The excited states of $N=44$ $^{74}$Zn were investigated via $\gamma$-ray spectroscopy following $^{74}$Cu $\beta$ decay. By exploiting $\gamma$-$\gamma$ angular correlation analysis, %with the GRIFFIN $\gamma$-ray spectrometer, 
the $2_2^+$, $3_1^+$, $0_2^+$ and $2_3^+$ states in $^{74}$Zn were firmly established.  %identified. 
The $\gamma$-ray branching and $E2/M1$ mixing ratios for transitions de-exciting the $2_2^+$, $3_1^+$ and $2_3^+$ states were measured, allowing for the extraction of relative $B(E2)$ values. In particular, the $2_3^+ \to 0_2^+$ and $2_3^+ \to 4_1^+$ transitions were observed for the first time. %The levels observed were organized into rotational-like bands 
%are identified in $^{74}$Zn at low energy, 
%and the results compared with large-scale shell-model calculations from which %provide 
%the shapes of individual states were determined.  %of the isotope in each of its states. 
The results show excellent agreement with new microscopic large-scale shell-model calculations, and are discussed in terms of underlying shapes, as well as the role of neutron excitations across the $N=40$ gap.
Enhanced axial shape asymmetry (triaxiality) is suggested to characterize $^{74}$Zn in its ground state. Furthermore, an excited $K=0$ band with a significantly larger softness in its shape is identified. A shore of the $N=40$ ``island of inversion'' appears to manifest above $Z=26$, previously thought as its northern limit in the chart of the nuclides.

\end{abstract}

\maketitle

The atomic nucleus can 
possess states at low excitation energy that have different shapes than the ground state, which is referred to as shape coexistence~\cite{garrett-2022,heyde-2011}. While this phenomenon seems to be ubiquitous, its most striking manifestations tend to appear in nuclei that have neutron or proton numbers corresponding to shell and subshell closures. Here, the energy gained through correlations can sometimes offset the spherical mean-field shell gaps, leading to the appearance of deformed low-energy ``intruder'' states in the ``normal'', near spherical, structure of the nucleus. In certain regions of the nuclear chart, referred to as ``islands of inversion''  (IOIs), the intruder configurations descend in excitation energy below the normal ones, thus becoming the ground states.  Understanding the ordering of the configurations, i.e., mapping their relative energies, permits tests of theoretical calculations of correlation energies. Currently, four IOIs are experimentally confirmed, associated with the neutron shell closures $N=8,~20,~28,\rm{and}~40$~\cite{Nowacki_PPNP}.%, and establishing their borders is an ongoing task for nuclear structure investigation.

Configurations leading to distinct shapes are often discussed in terms of axially symmetric prolate or oblate shapes coexisting with spherical states \cite{garrett-2022,heyde-2011}, but may also involve deviations from axial symmetry. The effects of nonaxiality have been observed for rapidly rotating nuclei (see, e.g., Refs.~\cite{hartley,odegard,petrache}), but there is less experimental evidence of its role in low angular momentum states near the ground state. 
In this respect, the most extensive information was obtained for $^{76}$Ge ($N=44$), for which the shape invariants deduced from a Coulomb-excitation study~\cite{ayangeakaa-2019} pointed to a rigid triaxial character.  Non-rigid triaxial structures can also occur,  where the nucleus may be imagined as fluctuating between  prolate and oblate shapes. This distinct structural paradigm is referred to as ``$\gamma$ softness''.  %Among the best examples of $\gamma$-soft nuclei are the Pt isotopes, with $^{194}$Pt explored in detail through Coulomb-excitation studies \cite{Wu-OsCoulex}. %The triaxial degree of freedom is also vital for nuclei exhibiting more diffuse shapes, and, 
The importance of the triaxial degree of freedom in the 
%for example, was shown to be required for the theoretical description of 
theoretical description of nuclei exhibiting more diffuse shapes was, for example, demonstrated for
$^{74,76}$Kr~\cite{clement,GIROD2009,bender2006}.
 %Perhaps the most subtle form of coexistence, however, occurs for situations where distinct {\it configurations} coexist -- we imply herein that one configuration possesses enhanced particle-hole correlations with respect to the other -- but for which the shapes may not be completely unique.  
The diffuseness of nuclear shapes may also conceal perhaps the most subtle form of coexistence, namely that of distinct {\it configurations} -- we imply herein that one configuration possesses enhanced particle-hole correlations with respect to the other -- but for which the shapes may not be completely unique.  

The northern border of the IOI at $N=40$ has so far been thought to occur in the $Z=26$ Fe nuclei, which for $36\le N\le 46$ present  a continuous decrease 
of the 2$^+_1$ excitation energy and an increase of the corresponding $B(E2)$ values \cite{rother2011}, pointing to deformation of their ground states.  Moreover, a significant occupancy of the neutron $g_{9/2}$ and $d_{5/2}$ orbitals, which appear above the energy gap for $N=40$, was necessary to reproduce the measured transition probabilities in $^{64,66}$Fe~\cite{ljungvall2010,klintefjord2017,olaizola2017JPG}. The ground states of magic $Z=28$ Ni nuclei are dominated by normal-ordered $0p0h$ configurations, while a multitude of low-lying 0$^+$ states were identified in $^{64-70}$Ni  \cite{nicu2020,leoni2017,olaizola2017,stryjczyk2018,chiara2012,BERNAS1982,chiara2015,prokop2015,MORALES2017,CRIDER2016}. Based on their decay properties combined with transfer-reaction cross sections \cite{darcey1971,flavigny2017}, these excited 0$^+$ states were interpreted as resulting from either neutron promotion across the energy gap for $N=40$ or proton excitation across the energy gap for $Z=28$, and tentatively assigned as intruder structures with various shapes.

The experimental information on the development of deformation and shape coexistence in the Zn nuclei ($Z=30$) is  more limited. 
Recently, on the basis of Coulomb-excitation measurements combined with large-scale shell-model (LSSM) and beyond-mean-field calculations, triaxiality of the ground states in $^{66,70}$Zn was proposed~\cite{rocchini-2021,calinescu-2021}. %which involved a certain degree of softness. 
The large quadrupole moment of the 5/2$^+$ isomer in $^{73}$Zn \cite{wraith} was also linked to triaxiality following guidance from Monte-Carlo shell-model  (MCSM) calculations that furthermore predicted considerable $\beta$ and $\gamma$ softness of the ground-state bands in $^{72,74}$Zn \cite{yang2018}. 
Regarding shape coexistence, $E0$ measurements~\cite{passoja-1985} hinted at the intruder character of the $0^+_2$ states in $^{64,66,68}$Zn, which for $^{66,68}$Zn was further supported by multi-step Coulomb-excitation data~\cite{rocchini-2021,koizumi-2004}. However, only for $^{68}$Zn was it possible to firmly assign different shapes to the $0^+_{1,2}$ states.  On the other hand,  shape coexistence in $^{79}$Zn was established through the observation of a large isomer shift for the 1/2$^+$ isomer~\cite{yang,orlandi}, %which based on the measured $g$ factor has been 
related to $2p1h$ neutron excitation across the $N=50$ shell gap. A sequence of non-yrast states in $^{78}$Ni was interpreted as belonging to a deformed intruder configuration~\cite{taniuchi-2019}, in line with the predictions of LSSM and MCSM calculations \cite{taniuchi-2019,nowacki-2016}. The presence of these deformed configurations was linked to the appearance of a new IOI at $N=50$ \cite{nowacki-2016}, which was predicted to merge with that at $N=40$ for nuclei with $Z \le 26$. In this context, it is pertinent to track how collectivity evolves across the Zn isotopic chain beyond $N=40$.
\begin{figure}[b!!]
   \includegraphics[width=\columnwidth]{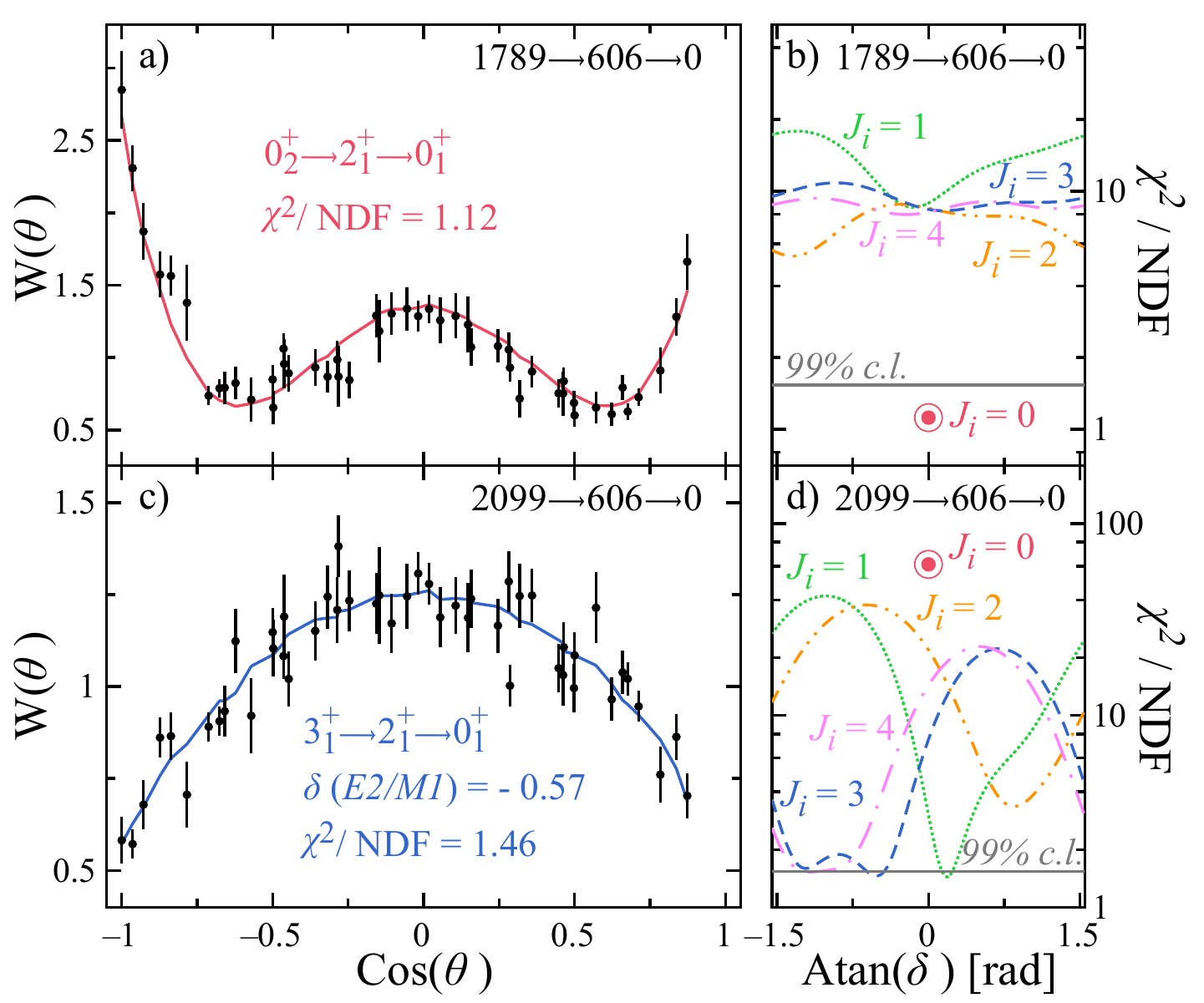}
   \caption{\label{fig:AC_results} %Measured angular correlation function $W(\theta)$ and results of the $\gamma-\gamma$ angular correlation analysis for the  $0_2^+ \to 2_1^+ \to 0_1^+$ (panels a and b) and $3_1^+ \to 2_1^+ \to 0_1^+$ cascades (panels c and d). Panels a and c show the experimental $W(\theta)$ functions, where $\theta$ is the opening angle between the GRIFFIN detectors. Panels b and d show the reduced $\chi^2$ as a function of the arctangent of the $E2/M1$ mixing ratio $\delta$. 
   Measured $\gamma$-$\gamma$ angular correlation functions $W(\theta)$, where $\theta$ is the opening angle between the GRIFFIN detectors, and reduced $\chi^2$ as a function of the arctangent of the  mixing ratio $\delta$, for the  $0_2^+ \to 2_1^+ \to 0_1^+$ (panels a and b) and $3_1^+ \to 2_1^+ \to 0_1^+$ cascades (panels c and d).
   The energies of the states involved in each cascade (in keV) are indicated. %as $E_i \to E_j \to E_f$.
   The $\chi^2$ distributions corresponding to 
   different spin hypotheses for the %higher-lying state with energy $E_i$ are labelled as in the figures. 
   initial state are labelled accordingly, with the continuous lines indicating the $99\%$ confidence limit. %is indicated with a dashed line in each panel.
   }
\end{figure}

In this Letter, we report information on the $^{74}$Zn excited states that combined with new LSSM calculations enable us to suggest: i) the IOI at $N=40$ extends above $Z=28$, and ii) configuration-coexisting structures possessing similar mean values of $\beta$ and $\gamma$, but that have significantly different degrees of softness, exist in the 
neutron-rich Zn isotopes.  The results presented herein rest on combining two key ingredients: the ability of the GRIFFIN spectrometer to perform $\gamma$-$\gamma$ angular correlation measurements with low beam intensities, and advancements with LSSM calculations that permit determination of shapes for specific states. 

The excited states in $^{74}$Zn were populated following $\beta$ decay of $^{74}$Cu produced at the TRIUMF-ISAC1 facility~\cite{dilling-2014} by spallation reactions of a 490-MeV proton beam impinging on a Ta target. The reaction products were ionized  using the TRIUMF Resonant Ionization Laser Ion Source (TRILIS)~\cite{bricault-2014} and then mass separated. The 25-keV $^{74}$Cu ions (at a rate of $1.7\times10^3$ s$^{-1}$) were implanted for about 40~h into a moving tape system positioned at the center of the GRIFFIN $\gamma$-ray spectrometer~\cite{garnsworthy-2019} equipped with 12 Compton-suppressed HPGe clover detectors. Decay data were obtained while the ions were collected on tape for 8~s, corresponding to about 5 half-lives of $^{74}\rm{Cu}$ [$1.63(5)$~s], and further observed for 1~s without the beam after which the tape was moved and the cycle repeated.  %, the tape was moved to limit the activity of the %$\beta$ decay of $^{74}$Zn [$T_{1/2}=95.6(12)$~s for the $0^+$ ground state] and $^{74}$Ga [$T_{1/2}=8.12(12)$~m for the $(3^-)$ ground state and $T_{1/2}=9.5(10)$~s for the $(0)$ isomer]. 
%longer-lived $^{74}$Zn and $^{74}$Ga daughters.
The standard GRIFFIN pre-sorting and data-correction procedures \cite{garnsworthy-2019} (e.g., summing and cross-talk corrections) were implemented in the analysis.

\begin{table*}
   \caption{\label{tab:results} Energies $E_\gamma$, branching ratios $I_\gamma$, mixing ratios $\delta(E2/M1)$ and relative $B(E2)$ values $B_{rel}(E2)$ measured in the present work, %are shown in columns 3, 5 and 6, respectively. Column 4 shows 
  together with branching ratios from Ref.~\cite{tracy-2018}. %Columns 7 and 8 show 
  Relative and absolute $B(E2)$ values % (the latter in Weisskopf units) 
  obtained from the present LSSM calculations %reported in the present work 
  (full diagonalization) are also given. %For the relative $B(E2)$ values, the transition used as a normalization for each state is indicated as  \textit{1}.
  Relative $B(E2)$ values of 100 are assumed for normalising transitions.}
   \begin{ruledtabular}
      \begin{tabular}{rrrrrrrr}
      $J_i^\pi \to J_f^\pi$   & $E_\gamma$~[keV] &   $I_\gamma$  &   $I_\gamma^{prev}$~\cite{tracy-2018}   &   $\delta(E2/M1)$ &      $B_{rel}^{exp}(E2)$   &   $B_{rel}^{SM}(E2)$    &   $B_{abs}^{SM}(E2)$~[W.u.]   \\
      \colrule
      $2_2^+ \to 2_1^+$ & 1064.32(10)  &   100.0(12) &   100.0(6)      &    $-1.13(6)$                           &   100(5)					&   100          &   9.7   \\
      $2_2^+ \to 0_1^+$ & 1670.07(20)  &   49.3(10)  &   49.4(4)       &                                         &   9.24(19)				&   22           &   2.1   \\
      $3_1^+ \to 2_2^+$ & 428.73(18)   &   6.5(4)    &   9.3(4)        &	$-0.8^{+0.2}_{-1.5}$                 &   $100^{+120}_{-30}$		&   100          &   40    \\
      $3_1^+ \to 4_1^+$ & 680.75(15)   &   7.10(19)  &   10.5(4)       &	$-1.0^{+0.3}_{-0.8}$                 &   $14^{+7}_{-5}$			&   7.8          &   3.1   \\
      $3_1^+ \to 2_1^+$ & 1493.2(3)    &   100.0(18) &   100.0(11)     &	$-0.57^{+0.06}_{-0.07}$              &   $1.9^{+0.4}_{-0.3}$    &   8.8          &   3.5   \\
      					&		       &    		 &				   &	$-2.7(5)$\footnote{Second solution.} &   6.8(4)					&				 &		  \\      
      $2_3^+ \to 0_2^+$ & 359.2(6)     &   2.0(4)    &                 &                                         &	 100(20)				&   100          &   17    \\
      $2_3^+ \to 2_2^+$ & 478.13(15)   &   6.8(7)    &   6.5(10)       &	$+0.9^{+0.8}_{-0.3}$                 &   $37^{+24}_{-15}$		&   15           &   2.6   \\
      $2_3^+ \to 4_1^+$ & 729.94(19)   &   3.1(7)    &                 &                                         &   4.5(10)				&   2.4          &   0.4   \\
      $2_3^+ \to 2_1^+$ & 1542.5(3)    &   37(3)     &   29.4(14)      &	$+2.4^{+1.8}_{-1.0}$                 &   $1.09^{+0.15}_{-0.26}$	&   0.18         &   0.03  \\
      $2_3^+ \to 0_1^+$ & 2148.73(16)  &   100(8)    &   100.0(27)     &                                         &   0.66(5)				&   0.18         &   0.03  \\
      \end{tabular}  
   \end{ruledtabular}
\end{table*}

The states and transitions below $\approx 3.1$~MeV observed in the recent $\beta$-decay study of $^{74}$Zn~\cite{tracy-2018} have been confirmed.  Previously, aside from % via  $\gamma-\gamma$ coincidences. However, contrary to previous works that for states other than 
the $2_1^+$ and $4_1^+$ states, only tentative spin assignments based on $\log(ft)$ values and model considerations were proposed.  
In the present work, definitive spin assignments from $\gamma$-$\gamma$ angular correlation analyses were made following the method described in Ref.~\cite{smith-2019}, based on the non-linear least-square fit to the measured correlation function $W(\theta)$ with the mixing ratio $\delta$ as a fit parameter~\cite{robinson-1990}. The finite size of the GRIFFIN detectors was accounted for by means of detailed GEANT4~\cite{agostinelli-2003} simulations. Following the recommendation of Ref.~\cite{robinson-1990}, only spin assignments for which $\chi^2$ results in a confidence level above $99\%$ were considered as definitive. The errors on the mixing ratios were evaluated by applying the $\chi_{min}^2+1$ condition ($68\%$ confidence level).

\begin{figure}
   \includegraphics[width=\columnwidth]{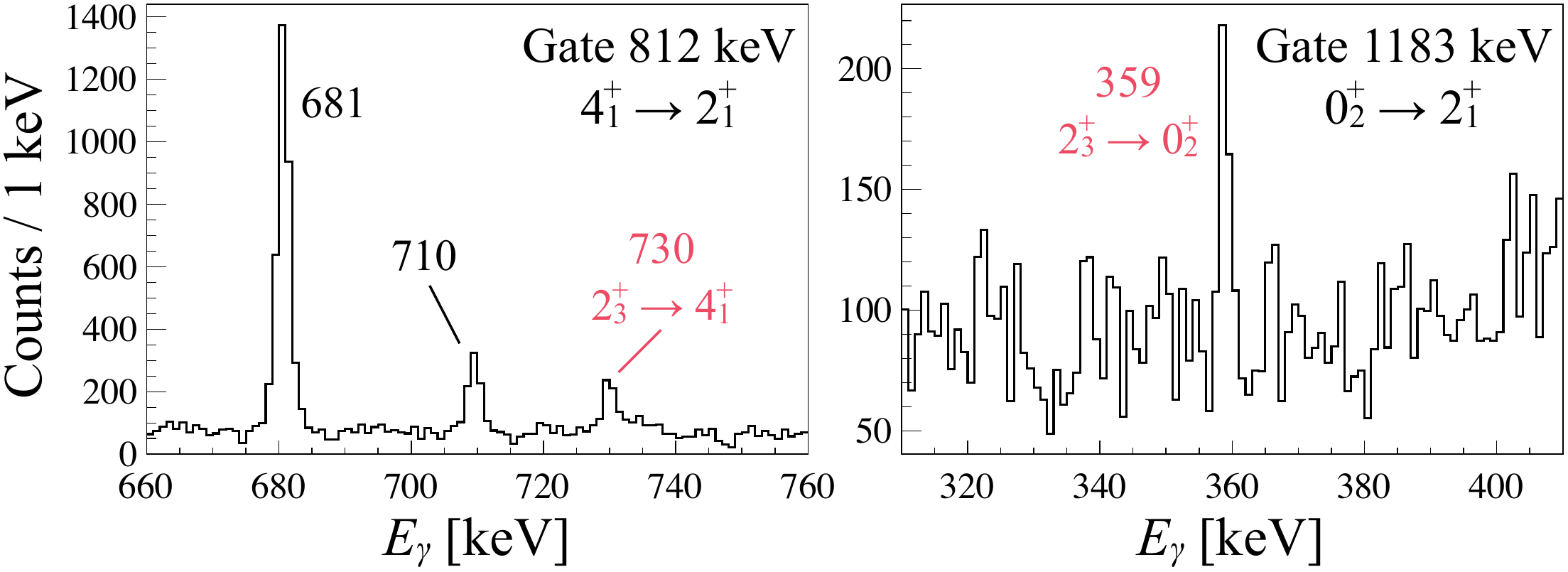}
   \caption{\label{fig:new_transitions}  Portions of the $^{74}$Zn $\gamma$-ray spectra in coincidence with $\gamma$ rays: (a)
   812 keV ($4_1^+ \to 2_1^+$) (b) 
   1183 keV ($0_2^+ \to 2_1^+$). The %729.94(19)-keV 
   730-keV and 359-keV %359.2(6)-keV 
   $\gamma$ rays are newly assigned as the $2_3^+ \to 4_1^+$ and $2_3^+ \to 0_2^+$ transitions, respectively. The 681-keV and 710-keV $\gamma$ rays were observed previously~\cite{tracy-2018}. %The 680.75(5)-keV and 709.69(13)-keV $\gamma$ rays have been observed previously~\cite{tracy-2018} in coincidence with the 812-keV one.
   %other known transitions in coincidence with the 812-keV one~\cite{tracy-2018}.
  }
\end{figure}

\begin{figure}
   \includegraphics[width=\columnwidth]{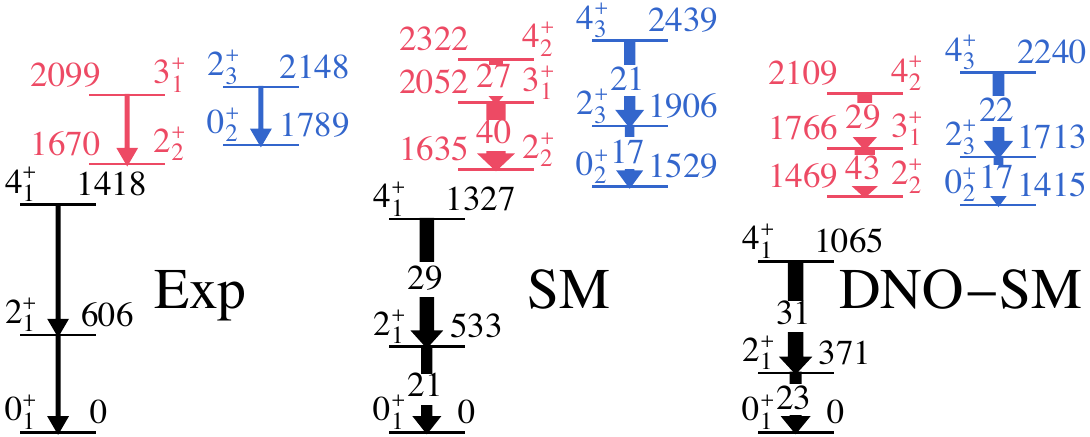}
   \caption{\label{fig:spectra} Partial experimental level scheme of $^{74}$Zn (EXP) compared with shell-model calculations (SM) and shell-model calculations in a deformed Hartree-Fock basis (DNO-SM). The states are labelled with their spin, parity, and energy (in keV) and organized in bands. Only in-band transitions are displayed, and their labels correspond to calculated $B(E2)$ values in W.u.}
\end{figure}

Examples of the $\gamma$-$\gamma$ angular correlations %and their analyses 
are shown in Fig.~\ref{fig:AC_results}. 
%With reference to Fig.~\ref{fig:spectra}, 
The 1789-keV state is firmly assigned as the  $0^+_2$ state with a unique solution (Fig.~\ref{fig:AC_results} panels a and b). Three different cascades for the 2099-keV state were analyzed, considering transitions to the $2_1^+$, $4_1^+$ and $2_2^+$ states. For the cascade involving the $2_1^+$ state, the $J=1,3,4$ assignments are possible (Fig.~\ref{fig:AC_results} panels c and d). However, since the 2099-keV state decays to the $4_1^+$ state, $J=1$ can be excluded. %Also, for $J=4$ one would expect a pure $E2$ transition to the $2_1^+$ state, which is clearly not the case.
%The fit obtained for $J=4$ requires a significant mixing for a 4->2 transition, which is unphysical for the E2/M3 case and thus that solution can be discarded.
%The $J=4$ solution can be discarded because it would require an unphysical $M3$ admixture.
If the $J=4$ solution was adopted, it would require a highly mixed transition of $M3/E2$ multipolarity, as shown in Fig.~1, for the decay to the first excited state. This large admixture is unphysical since it would lead to a highly enhanced $B(M3)$ value, and thus this solution can be discarded and the 2099-keV state is firmly identified as the $3^+_1$ state. 
The 1670-keV and 2148-keV states are assigned as $J=2$ with unique solutions and identified as the $2_2^+$ and $2_3^+$ states, respectively. 
Additionally, two  
transitions were observed for the first time in the present work: $2_3^+ \to 0_2^+$ at %359.2(6)~keV 
359~keV and $2_3^+ \to 4_1^+$ at %729.94(19)~keV 
730~keV as shown in Fig.~\ref{fig:new_transitions}. These sequences of levels and spins are suggestive of excited $K=0$ and $K=2$ structures. 
Using the branching and mixing ratios, relative $B(E2)$ values were determined (Tab.~\ref{tab:results}). 
The strong relative $B(E2;2_3^+ \to 0_2^+)$ and $B(E2;3_1^+ \to 2_2^+)$ values support the assignment of the $2^+_3$ and $3^+_1$ levels as rotational band members built on the $0_2^+$ and $2_2^+$ states, respectively.  These key experimental results are displayed in Fig.~\ref{fig:spectra}. 

The structure of $^{74}$Zn was further investigated within the shell-model framework. The large valence space employed comprised the $pf$ shell for protons and the $1p_{3/2}0f_{5/2}1p_{1/2}0g_{9/2}1d_{5/2}$ orbitals for neutrons, and thus incorporated the  degrees of freedom required for the description of collectivity at the $N=40$ interface and the breaking of the $Z=28$ and $N=40$ cores. The LNPS effective interaction \cite{lenzi-2010} was used, with recent minor adjustments to extend its reliability up to $N=50$ and account for particle-hole excitations \cite{nowacki-2016,taniuchi-2019}.

\begin{figure}[b!!]
   \includegraphics[width=\columnwidth]{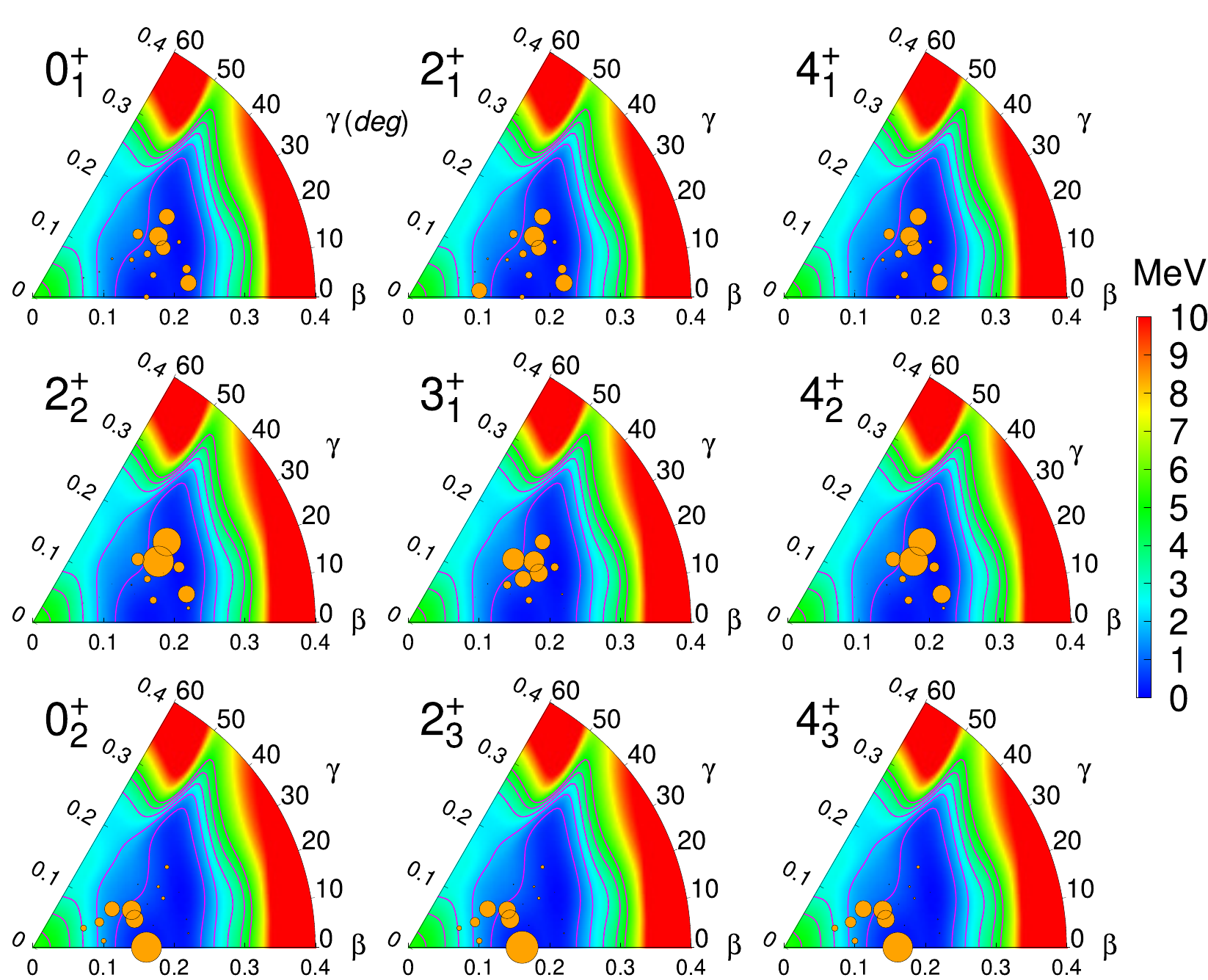}
   \caption{\label{fig:shapes} Normalized probability to find a deformation $(\beta,\gamma)$ in specific $^{74}$Zn states represented with circles on the %potential energy surface. 
   PES, whose radii are proportional to the probability.}
\end{figure}

Firstly, using the shell-model framework with the same valence space and effective hamiltonian, the potential energy surface (PES) of $^{74}$Zn was obtained from constrained Hartree–Fock calculations  (see Fig.~\ref{fig:shapes}). At the mean-field level, $^{74}$Zn exhibits a non-spherical minimum with $\beta \approx 0.2$, extending towards a %non-vanishing 
triaxial shape (similar conclusions were reached from the PES calculated using the Gogny D1S interaction \cite{illana2014}). To go beyond the mean-field level, by mixing 
the deformed Hartree-Fock solutions through the generator coordinate method (dubbed as DNO-SM in~\cite{dao-tbp}),   the level scheme and in-band $B(E2)$ values, presented in Fig.~\ref{fig:spectra}, are obtained and agree well with the experimental values (a slight compression of the level scheme with respect to the results of the full SM calculation comes from the DNO-SM basis truncation). Three
bands with large in-band $B(E2)$ values emerge: a rotational ground-state band (g.s.b.), a `$\gamma$' band related to it, and a third band built on the $0^+_2$ state. %The extracted normalized probability amplitudes of specific $K$ components in each $J \neq 0$ state show that the $K=0$ ground-state, $K=2$ `$\gamma$', and $K=0$ $0_2^+$ bands have small (below 1\%) admixtures of other $K$ components. Their appearance is in line with a non-axial character of  $^{74}$Zn.
For $J \neq 0$ states in the $K=0$ ground-state, $K=2$ `$\gamma$', and $K=0$ $0_2^+$ bands, small (below 1\%) admixtures of other $K$ components were found, in line  with a non-axial character of  $^{74}$Zn.

Figure~\ref{fig:spectra} also presents the $^{74}$Zn level scheme  calculated using the full shell-model diagonalization. The agreement with the experiment is excellent for all considered states. 
The relative $B(E2)$ values resulting from this approach, shown in Tab.~\ref{tab:results}, demonstrate that the calculation correctly predicts the dominant decay paths of 
the $2_2^+$, $3_1^+$ and $2_3^+$ states. The  $B(E2;3_1^+ \to 2_2^+)/B(E2;3_1^+ \to 4_1^+)$  and $B(E2;2_3^+ \to 0_2^+)/B(E2;2_3^+ \to 2_2^+)$ ratios are also reproduced within the uncertainties. Differences of a factor of $2-3$ in weak transitions are observed. These transitions, however, arise from overlaps of small components in the wave functions. From the in-band $B(E2)$ values, 
summarized in Fig.~\ref{fig:spectra}, $\beta=0.23$ for the ground-state and 
`$\gamma$' bands were calculated (in agreement with the PES), and a lower 
$\beta=0.20$ for the band built on the $0_2^+$ state. 
The hindrance of transition probabilities calculated between the band built on the $0_2^+$ state and g.s.b.\ ($B(E2;0_2^+ \to 2_1^+)=0.58$~W.u.\ and $B(E2;2_3^+ \to 0_1^+)=0.03$~W.u.) excludes a $\beta$-vibrational origin of the former \cite{Garrett_beta}, and is instead compatible with a configuration-coexistence scenario with weak mixing.

Further details regarding the shapes of specific states were obtained within two different approaches. 
Figure~\ref{fig:shapes} presents the normalized probability to find a specific $(\beta, \gamma)$  deformation in each state.
Alternatively, from the Kumar quadrupole sum rules~\cite{kumar-1972,cline-1986,poves-2020}, $\langle \beta \rangle = 0.24$ and $\langle \gamma \rangle = 24^\circ$ for the ground state and $\langle \beta \rangle = 0.22$, $\langle \gamma \rangle = 20^\circ$ for the $0_2^+$ state were obtained.  
The results of these two procedures are consistent, and confirm the picture of a triaxially deformed ground state coexisting with a band built on the $0_2^+$ 
state,  which is slightly less deformed than the ground state but exhibits an 
extended softness in the $\gamma$ degree of freedom towards the axial prolate 
shape.  
The fluctuations $\sigma$ in both $\beta$ and $\gamma$~\cite{poves-2020} are important for both states: $\sigma_{\beta} (0_1^+) = \pm 0.04$, $\sigma_{\gamma} (0_1^+) = (+11^\circ, - 13^\circ)$, $\sigma_{\beta} (0_2^+) = \pm 0.04$, $\sigma_{\gamma} (0_2^+) = (+12^\circ, - 20^\circ)$.  %Referring to Fig.~\ref{fig:shapes}, we see the origin of the much larger value for the variance in $\gamma(0^+_2)$; there are components in the wave functions for the $0^+_2$ band members with a large weight that lie on the prolate axis that are absent for the g.s.b.  
%The origin of a significantly larger value for the variance in $\gamma(0^+_2)$ becomes clear when looking at Fig.~\ref{fig:shapes}: the wave functions for the $0^+_2$ band members involve components, with a large weight, which lie on the prolate axis; these are absent for the g.s.b.
The difference between  $\sigma_{\gamma}$ values for the $0^+_1$ and $0^+_2$ states can be explained by the presence of important components in the $0^+_2$ band members' wave functions, which lie on the prolate axis; these are absent for the g.s.b. (Fig.~\ref{fig:shapes}).
Even though the mean $\beta$ and $\gamma$ values  extracted from the sum rules are very similar for both states, the underlying distributions in the $(\beta,\gamma)$ plane are substantially different, leading us to assert that the configurations represent different, but overlapping, shapes.  Interestingly, the $\sigma_{\beta}$ and $\sigma_{\gamma}$ values calculated for the ground states in $^{74}$Zn and in 
its triaxial isotone $^{76}$Ge~\cite{poves-2020} are very similar. 

\begin{figure}[t]
   \includegraphics[width=\columnwidth]{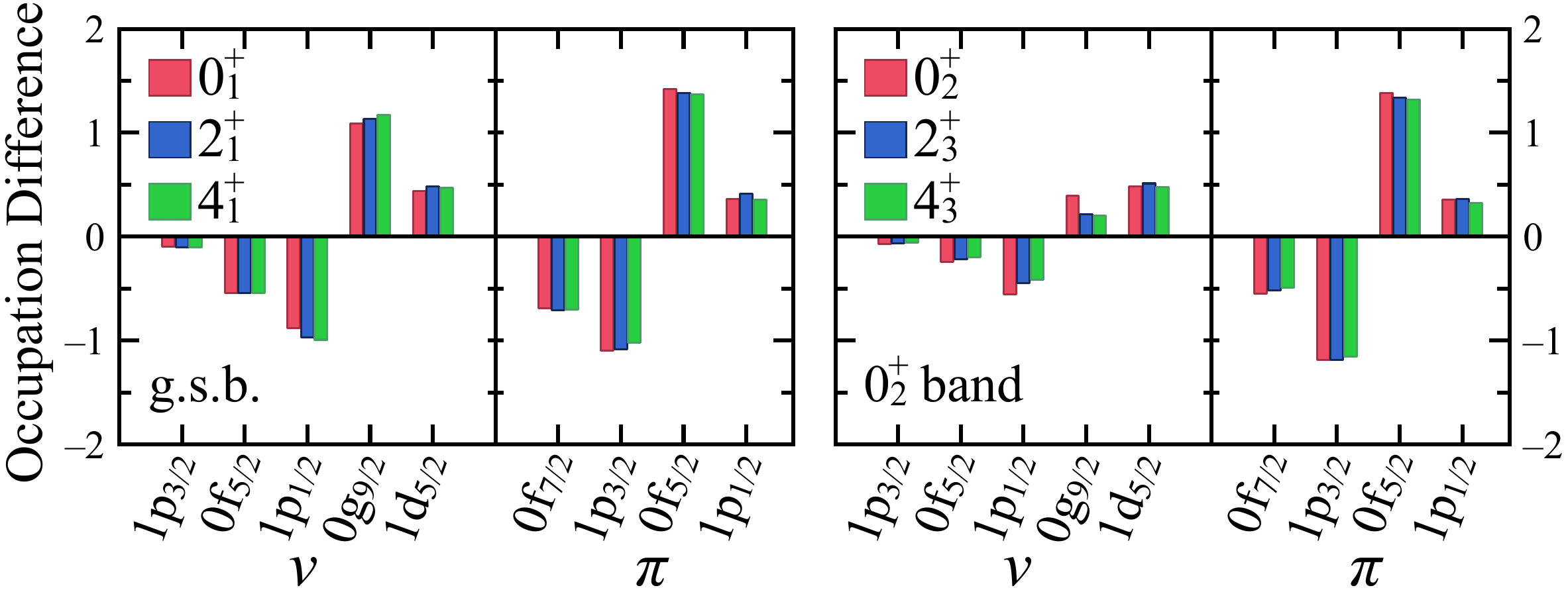}
   \caption{\label{fig:occupations} Difference in occupation numbers with respect to the normal filling of the proton ($\pi$) and neutron ($\nu$) orbitals considered in the present LSSM calculations for $^{74}$Zn (left panel: ground-state band, right panel: band built on the $0_2^+$ state). %The $^{74}$Zn ground-state band is shown on the left while the band built on the $0_2^+$ state on the right. 
   %Neutron and proton orbitals are denoted with $\nu$ and $\pi$, respectively.
   }
\end{figure}

\begin{figure}
   \includegraphics[width=\columnwidth]{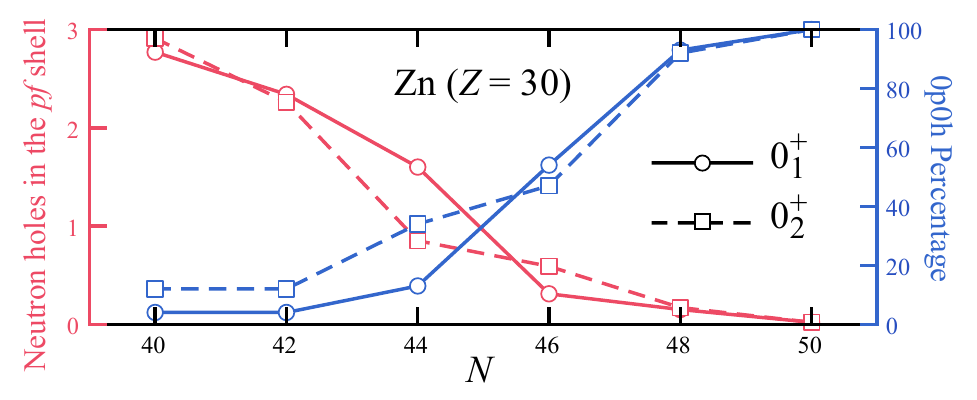}
   \caption{\label{fig:excitations} Neutron excitations from the $pf$ shell into the $0g_{9/2}$, $1d_{5/2}$ orbitals (red) and percentage of the $0p0h$ configuration (blue) in the ground state and the $0^+_2$ state (continuous and dashed lines, respectively) in Zn isotopes with $40\le N\le50$.}
\end{figure}

Figure~\ref{fig:occupations} shows the occupation numbers in the neutron and proton orbitals considered in the present LSSM calculations for the $^{74}$Zn g.s.b.\ and the excited $K=0$ band, up to $J=4$. The wave-function compositions are consistent within each of the two bands, but significantly different when comparing the two. For the g.s.b., 
approximately two neutrons with respect to the normal filling are excited from the $pf$ shell across the energy gap for $N=40$. 
This can be contrasted with the average of less than one neutron excited for the $0_2^+$ band, while the occupation of proton orbitals is nearly identical in the two bands. 
As shown in Fig.~\ref{fig:excitations}, the number of neutrons excited from the 
$pf$ shell  for both the ground  and the $0^+_2$ states 
decreases from a maximum of three to zero between $N=40$ $^{70}$Zn and $N=50$ $^{80}$Zn, while the percentage of the $0p0h$ configuration increases from close to zero to $100\%$ over the same range. 
The structures of the $0_{1,2}^+$ states exhibit the largest difference for $^{74}$Zn, for which the contribution of the $0p0h$ configuration to the ground state is less than a half of that to the $0_{2}^+$ state.
A similar behaviour is
observed in the neighbouring $^{74,76}$Ge isotopes: %in which the  $0^+_2$ state is also interpreted as a shape-coexisting state, 
the $0_{2}^+$ states
are less deformed and with fewer neutron excitations across the energy gap for $N=40$ than the ground states~\cite{heyde-2011,honma-2009,toh-2001,toh-2000}. 
Moreover, the predominance of multiparticle-multihole configurations in the structure of the ground states of the Zn isotopes with $N<46$ suggests
that these nuclei %constitute the  ``shore'' of 
belong to the $N=40$ IOI, which extends beyond $Z=28$. Contrary to other IOI borders, this one is not reflected in a sudden change of ground-state properties.

To summarize, the present experimental and theoretical results provide evidence for an unexpected enhanced triaxial deformation of $^{74}$Zn. %Correct identification of regions of the nuclear chart where triaxiality plays a role is  relevant in the context of modelling of astrophysical processes, as it is known to significantly impact the nuclear masses~\cite{moller-2006,moller-2008} that are used as an input in such calculations. 
%While previous mean-field calculations~\cite{moller-2008,awwad-2020} predicted triaxiality in the neutron-rich Zn nuclei, it was expected to appear much farther from stability (i.e., starting from $^{95}$Zn~\cite{moller-2008}). 
This has implications beyond nuclear structure, as triaxiality is known to significantly impact the nuclear masses~\cite{moller-2006,moller-2008} that are used as an input when modelling  astrophysical processes.
Moreover, the identification of a coexisting $K=0$ band bridges the gap between the Ge and Ni isotopes, in which such structures are well established. 
The ground state of $^{74}$Zn is suggested to involve more neutron excitations across the $N=40$ gap than the $0^+_2$ state, which indicates that the $N=40$ IOI does not end at $Z=26$ as previously assumed, but extends further north in the nuclear chart. Future experimental work in this mass region should involve direct reaction studies to probe the microscopic content of the wave functions, combined with measurements of absolute quadrupole and monopole transition strengths.%Extracting absolute transition probabilities in this mass region is a future challenge for next-generation radioactive ion facilities. For $^{74}$Zn, the experimental results obtained in the present experiment will offer the baseline for Coulomb excitation and lifetime measurements.

\begin{acknowledgments}

The authors thank the operations and beam delivery staff at TRIUMF for providing the high-quality $^{74}$Cu radioactive beam. This work was supported by the Natural Sciences and Engineering Research Council of Canada. The GRIFFIN infrastructure was funded jointly by the Canada Foundation for Innovation, the Ontario Ministry of Research and Innovation, the British Columbia Knowledge Development Fund, TRIUMF and the University of Guelph. TRIUMF receives funding through a contribution agreement with the National Research Council of Canada.
This work was also supported by the U.S. Department of Energy (DOE) under contract no. DE-FG02-93ER40789. The TTU team was sponsored by the Office of Nuclear Physics, U.S. Department of Energy, under contract no. DE-SC0016988.

\end{acknowledgments}

%\bibliography{74Zn-GRIFFIN}

\providecommand{\noopsort}[1]{}\providecommand{\singleletter}[1]{#1}%

\end{document}